\documentclass[%
 amsmath,amssymb,
 aps,
 preprint,
 reprint,
floatfix,
]{revtex4-2}
\usepackage{placeins}
\usepackage{comment}
\usepackage[normalem]{ulem}
\usepackage{subfiles} 
\usepackage{color}
\usepackage{soul}
\usepackage{float}
\usepackage{placeins}
\usepackage[standard]{ntheorem}
\usepackage{graphicx}% Include figure files
\usepackage{dcolumn}% Align table columns on decimal point
\usepackage{bm}% bold math
\usepackage{graphics}
\usepackage{graphicx}
\usepackage[unicode, bookmarks, colorlinks, breaklinks]{hyperref}  
\newcommand*{\QEDA}{\null\nobreak\hfill\ensuremath{\square}}

\newcommand{\norm}[1]{\left\lVert#1\right\rVert}

\newcommand{\tbold}[1]{\tilde{\mathbf{#1}}}

\newcommand{\cv}[1]{\mathbf{#1}}
\newcommand{\cs}[1]{\boldsymbol{#1}}

\newcommand{\bs}[1]{\boldsymbol{#1}}
\newcommand{\bb}[1]{\mathbf{#1}}
\newcommand{\crl}[1]{{\color{blue} #1}} % Carlo's comments
\newcommand{\alb}[1]{{\color{green} #1}} %  Alberto's comments
\usepackage{cancel}
\usepackage{subfiles}

\begin{document}

\preprint{APS/123-QED}

\title{Optimal Control of Short-Time Attractors in Active Fluid Flows}% Force line breaks with \\

\author{Carlo Sinigaglia}
%\email{carlo.sinigaglia@polimi.it}
\author{Francesco Braghin}%
% \email{francesco.braghin@polimi.it}
\affiliation{%
 Politecnico di Milano, Department of Mechanical Engineering, 20156, Italy
}%


\author{Mattia Serra}
 \email{mserra@ucsd.edu}
\affiliation{University of California San Diego, Department of Physics, CA 92093, USA
}%
%\affiliation{
%\affiliation{%


\date{\today}% It is always \today, today,

\begin{abstract}
%Short-time attractors are closely related to the instantaneous limit of the well-known finite time Lyapunov exponent which, up to first-order, can be described by the minimum eigenvalue of the rate of strain tensor. In this paper, we propose an optimal control method to generate short-time attractors in active fluid flows. The active fluid dynamics is described by an elliptic partial differential equation where the control action enters linearly through a linear differential operator which model the active stress mechanism. We formulate an optimal control problem to track a desired profile of the minimum eigenvalue of the rate of strain tensor relative to the state dynamics. By properly designing the target function, we generate short-time attractors from an Eulerian perspective. A system of first order necessary optimality conditions is derived following a standard Lagrangian method. The forcing term in the adjoint dynamics is derived from the first variation of a cost functional involving the minimum eigenvalue of the rate of strain tensor. Numerical simulations show the effectiveness of the proposed approach in tracking an arbitrary profile of the minimum eigenvalue of the rate of strain tensor while rejecting disturbances.

Objective Eulerian Coherent Structures (OECSs) and instantaneous Lyapunov exponents (iLEs) govern short-term material transport in fluid flows as Lagrangian Coherent Structures and the Finite-Time Lyapunov Exponent do over longer times. Attracting OECSs and iLEs reveal short-time attractors and are computable from the Eulerian rate-of-strain tensor. Here we devise an optimal control strategy to create short-time attractors in viscosity-dominated active fluids. By modulating the active stress intensity, our framework achieves a target profile of the minimum eigenvalue of the rate-of-strain tensor, controlling the location and shape of short-time attractors. We use numerical simulations to show that our optimal control strategy effectively achieves desired short-time attractors while rejecting disturbances. Combining optimal control and recent advances on coherent structures, our work offers a new perspective to steer material transport in unsteady flows, with applications in synthetic active nematics and multicellular systems.

%\begin{description}
%\end{description}
\end{abstract}

\maketitle

Large-scale coherent dynamics where global collective behaviors arise from local interactions,  individual anisotropies and activity are ubiquitous. Bird flocks, bacterial swarms or ensembles of cells exhibit macroscopic patterns whose length scale is orders of magnitude larger than the individual size \cite{Toner1995,Vicsek2012,RevModPhys.85.1143,kruse2004asters,ballerini2008interaction,zhang2010collective,bricard2013emergence,dombrowski2004self,copenhagen2021topological,meacock2021bacteria,friedl2009collective,ladoux2017mechanobiology,Doostmohammadi2018,serra2020dynamic}. 
%The macroscopic dynamics of these large-scale systems of individuals, sometimes referred to as living or active matter (see e.g. the survey \cite{Marchetti2013}), sparked an increasing interest due to its unusual physical properties such as self-organization, broken symmetry, many-body phases and non-reciprocity which are not subject to standard laws of mechanics, see e.g. \cite{Fruchart2021}.
The macroscopic dynamics of these systems of active individuals --or active matter-- exhibit nonstandard physical properties such as self-organization, symmetry breaking and non-reciprocity \cite{ramaswamy2017active,Marchetti2013,toner2005hydrodynamics,Fruchart2021,bowick2022symmetry,shankar2022topological}. 
%These dynamic phases depend on the systems' energy intake and dissipation mechanisms. Indeed, the non-conservative active stresses break the symmetry that passive materials would possess, see e.g., \cite{shankar2022topological} for an analysis of robustness to perturbations in active systems using arguments from topology.
There are several descriptions of active matter systems \cite{shaebani2020computational}, including agent-based models, coarse-grained continuum models, and data-driven models \cite{joshi2022data}. Here we focus on hydrodynamic models, which predict the macroscopic behavior of the system from a small set of parameters and take the form of Partial Differential Equations (PDEs) describing quantities such as velocity, density and nematic tensor. 

Besides studying the emergent properties of active matter, it is natural to ask how one can control such systems to steer their global dynamics. The main possibilities rely on distributed or boundary control techniques \cite{MQS}. Experimentally, \cite{ross2019controlling} generated desired persistent fluid flows by regulating light patterns on a mixture of optogenetically modified motor proteins and microtubule filaments. Also controlling light, \cite{lemma2022spatiotemporal} achieved spatiotemporal patterning of extensile active stresses in microtubule-based active fluids. By controlling an external electric field affecting cellular signaling networks, \cite{cohen2014galvanotactic} steered the collective motion of MDCK-II epithelial cells. From a theoretical perspective, \cite{shankar2022spatiotemporal} propose a new framework to steer topological defects --the localized singularities in the orientation of the active building blocks \cite{RevModPhys.85.1143}-- by controlling activity stress patterns. \cite{norton2020optimal} devised an Optimal Control Problem (OCP) to achieve a target nematic director field by controlling either an applied vorticity field in the nematic tensor dynamics or the active stress magnitude in the velocity dynamics. Alternative control strategies use surface anchoring at the boundaries and substrate drag to rectify the coherent flow of an active polar fluid in a 2D channel \cite{boundaryControlGulati2022}. %\textcolor{red}{Carlo, puoi leggere l'intro di questa referenza recente e valida \cite{shankar2022spatiotemporal}, e vedere se stiamo mancando paper rilevanti su control of active matter? grazie!} \crl{Le reference che cita su optimal control sono le stesse che citiamo noi, poi lui utilizza un altro metodo basato sulla sovrapposizione di una sequenza di active stress intensity e argomenti di simmetria }

\begin{figure*}
\centering 
\includegraphics[width = 1\linewidth]{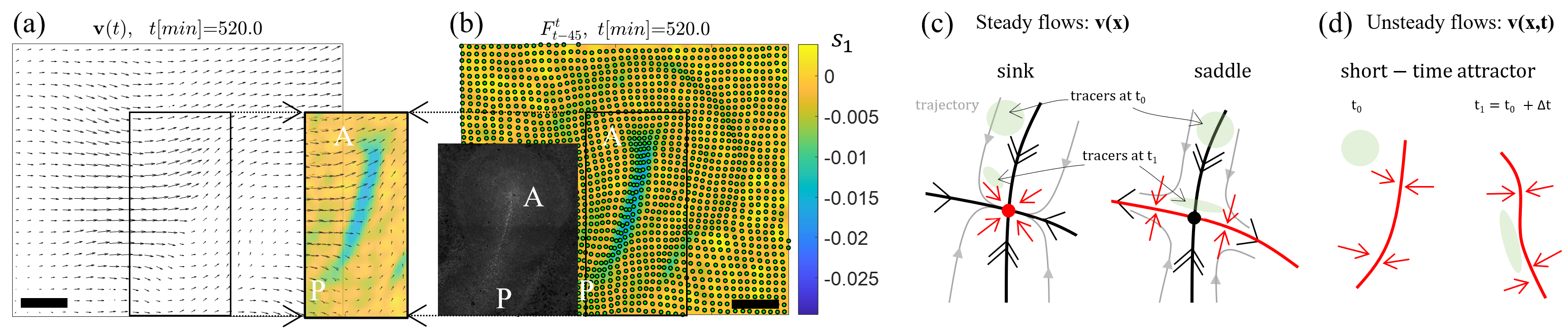}
\caption{
\label{fig:s1_vel_chick}
\textcolor{black}{Example of short-time attractors in multicellular flows. The chicken embryo's epiblast contains $\approx 60,000$  cells during gastrulation, which occurs over $\approx$ 12 hours. Velocities are reconstructed using particle image velocimetry and Light Sheet Microscopy \cite{Rozbicki2015}. The scale bar is $500 \mu m$, and $t = 0$ corresponds to the beginning of gastrulation. (a) Velocity field $\mathbf{v}(\mathbf{x},520)$ (black vectors) and (b) instantaneous Lyapunov Exponent field $s_1(\mathbf{x},520)$ (colormap), consisting of the smallest eigenvalue of the rate-of-strain tensor of $\mathbf{v}$. The left inset in (b) shows the corresponding fluorescence image, and AP denotes the Anterior-Posterior axis of the embryo. Negative values of $s_1(\mathbf{x},520)$ mark short-time attractors, which are undetected by inspection of $\mathbf{v}$ (insets in a). To visualize the effect of short-term attractors, green dots in panel b mark the current $t=520$ position $\mathbf{F}_{475}^{520}(\mathbf{x_0}) = \mathbf{x_0} + \int_{475}^{520} \mathbf{v}(\mathbf{F}_{475}^{\tau},\tau)d\tau$ of short-time trajectories of $\mathbf{v}(\mathbf{x},t)$, starting at $t=475$ from a uniform spatial configuration. Within this short time ($45\ min/12h \approx 6\%$ of gastrulation time), trajectories accumulate on the $s_1(\mathbf{x},520)$ trench. See Fig. S2 for the same analysis at a different time. (c) Examples of asymptotic attracting structures in a steady velocity field $\mathbf{v}(\mathbf{x})$ include a sink fixed point (red circle) and the unstable manifold (red curve) of a saddle fixed point (black circle). These attracting structures are stationary and anchored to a fixed point of $\mathbf{v}(\mathbf{x})$. In unsteady flows $\mathbf{v}(\mathbf{x},t)$, these notions are unsuitable for predicting material attraction: they produce false positives and false negatives \cite{serra2016objective}. By contrast, short-term attractors (d) provide the correct material attraction locations, which can move and deform: the velocity is nonzero and nonconstant along the short-term attractor (insets in panel a).}}
\end{figure*}

Existing theoretical methods target a desired configuration of the nematic director field, topological defects or fluid velocities. While defects' dynamics drive large-scale chaotic flow \cite{Giomi2015,PhysRevX.9.041047,Tan2019}, they may not be enough to predict spatiotemporal material transport. For instance, \cite{serra2021defect} shows in experimental and numerical active nematics that the director field alone cannot predict if different domain regions will mix over a desired time interval or remain separated by a transport barrier, as well as predict where transport barriers are. In fact, even the knowledge of the velocity field and typical streamline or vorticity plots are sub-optimal to studying material transport in unsteady flows, as shown in experimental and simulated velocities \cite{LCSHallerAnnRev2015,serra2016objective,serra2020SR} \textcolor{black}{and Figure \ref{fig:s1_vel_chick}}.

A natural framework to quantify material transport is the concept of Coherent Structures (CSs), see e.g. \cite{LCSHallerAnnRev2015,SerraHaller2015,hadjighasem2017critical}, which serve as the robust frame-invariant skeletons shaping complex trajectory patterns. CSs include Lagrangian Coherent Structures (LCSs)\cite{LCSHallerAnnRev2015} which organize material transport over a finite time interval, and their short-time limits called Objective Eulerian Coherent Structures (OECSs) \cite{SerraHaller2015,nolan2020finite}.  
Attractors, their domain of attraction and repellers are widespread CSs in embryonic development across species \cite{serra2020dynamic,serra2021mechanochemical,lange2023zebrahub} and active nematics \cite{serra2021defect}. Controlling material transport in active matter enables the modulation of spatial mixing or trapping in synthetic systems but also enhances our understanding and perhaps the ability to influence multicellular flows in embryonic development. 

Here, we consider the OCP of a simplified version of the compressible active fluid introduced in \cite{serra2021mechanochemical,chuai2023reconstruction} to describe multi-cellular movements in vertebrate gastrulation. In the hydrodynamic approximation, the system's states are described by PDEs that couple the multicellular fluid velocity $\cv{v}$, the orientational dynamics $\phi$, which dictates the anisotropy of the active stresses, and $m$ which describes the active stress intensity. We seek to control the location and shape of short-time attractors marking regions of material accumulation. 
 
\section{Material Transport}
\begin{figure*}[t]
\includegraphics[width = .9\linewidth]{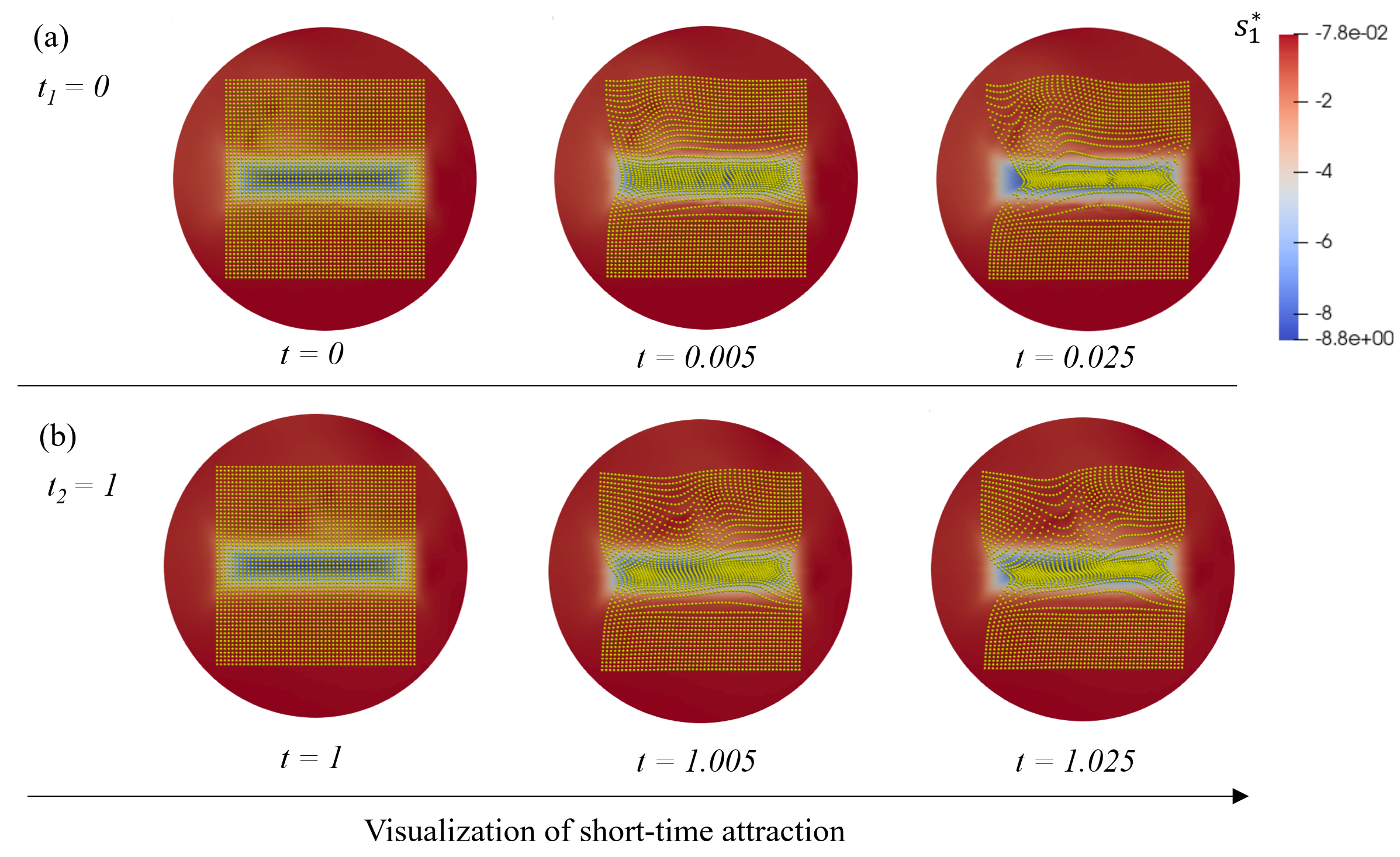}
\caption{
\label{st_attr}
Optimal solution of the nonlinear OCP generating short-time attractors. The target minimum eigenvalue $z(\bb{x})$ is the indicator function of a rectangle at the center of the domain. The optimal eigenvalue field $s_1^{\star}(\mathbf{x})$ is shown by the colormap. Each row (a-b) corresponds to a different initialization time ($t_i,\ i=1,2$). %characterized by a disturbance field $\mathbf{d}(\mathbf{x},t_i)$ which varies on a slow time scale compared to the short attraction time scales. 
In each row, we visualize the effect of short-time attractors by initializing a uniform set of fluid tracers (yellow dots) at each $t_i$ and displaying their later positions integrating $\mathbf{v}(\mathbf{x},t)$ over short times, as in Figs. \ref{fig:s1_vel_chick}b. 
}
\end{figure*}
 Long-term material transport is a fundamentally Lagrangian phenomenon, originally studied by keeping track of the longer-term redistribution of individual tracers released in the flow. In that setting, the Finite Time Lyapunov Exponent (FTLE) and Lagrangian coherent structures (LCSs) have been efficient predictors of tracer behavior see e.g. \cite{haller2015lagrangian,shadden2005definition,hadjighasem2017critical,serra2017uncovering,serra2020dynamic,serra2021defect}. An alternative
to Lagrangian approaches is to find their instantaneous limits purely from Eulerian observations, thereby avoiding  the pitfalls of trajectory integration, while predicting  short-term
material transport. Additionally, LCSs are generally impractical to control -- no literature exists -- because they are defined as nonlinear functions of fluid trajectories, which are themselves integrals of the Eulerian velocity $\cv{v}$. %By contrast, Eulerian structures are defined as functions of $\cv{v}$. 

Short-time attractors -- originally defined as Attracting OECSs \cite{serra2016objective}  --  govern material transport in fluid
flows over short-times, revealing critical information in challenging problems such as search and rescue operations at sea \cite{serra2020search} and oil-spill containment \cite{duran2021horizontal}. A simpler, more controllable alternative to attracting OECSs for locating short-time attractors is the instantaneous Lyapunov Exponent (iLE)\cite{Nolan2020}, defined as the instantaneous limit of the well known FTLE. The iLE locates short-time attractors as trenches -- or negative regions -- of the smallest eigenvalue $s_1$ of the rate-of-strain tensor of the fluid velocity $\cv{v}$. \textcolor{black}{For example, Figs.\ref{fig:s1_vel_chick}a-b show short-term attractors marked by trenches of $s_1$ (scalar field) in an experimental velocity field (black vector field) describing the motion of thousands of cells during chick gastrulation \cite{Rozbicki2015}. A strong trench of $s_1$ marks a short-term attractor along the anterior-posterior (AP) axis corresponding to the forming primitive streak \cite{serra2020dynamic} (panel b), while remaining not identifiable from the inspection of the corresponding velocity field (panel a and its inset). Similar results hold in different flows (see e.g., Fig. 1 of \cite{serra2016objective} and Figs. 4-5 of \cite{serra2020search}). To confirm the effect of short-term attractors, panel b shows that tracers (green dots) released from a uniform grid of initial conditions and advected by $\mathbf{v}(\mathbf{x},t)$ for a short time accumulate on the $s_1$ trench.} 

\textcolor{black}{Classical asymptotic attracting structures (Fig. \ref{fig:s1_vel_chick}c) include sink-type fixed points (red dot) and the unstable manifold (red curve) of saddle-type fixed points of steady velocity fields. These structures, however, cannot move or deform in time and influence material transport only for steady velocity fields.} 
\textcolor{black}{As shown in  Figs.\ref{fig:s1_vel_chick}a-b, inspection and control of the velocity field $\mathbf{v}$ is sub-optimal to create material traps in general unsteady flows. First, because the velocity field and its streamlines are not objective, i.e., they depend on the choice of reference frame used to describe motion (see also SM Section 6 and Figs. S1-S2). By contrast, the location of material accumulation is frame invariant \cite{haller2015lagrangian,serra2016objective}, as any quantification of the material response of a deforming continuum must be according to a fundamental axiom of mechanics \cite{gurt}. Second, it might be an unnecessarily strong requirement, or uncompliant with boundary conditions, to prescribe $\mathbf{v}(\mathbf{x},t)$ directly.} %\textcolor{red}{Classical asymptotic attracting structures (Fig. \ref{fig:s1_vel_chick}c) include sink-type fixed points (red dot) and the unstable manifold (red curve) of saddle-type fixed points of steady velocity fields. These structures, however, cannot move or deform in time, are frame-dependent, and influence material transport only for steady velocity fields.} %By contrast, short-time attractors (Fig. \ref{fig:s1_vel_chick}d) can move and deform, are frame invariant and shape material transport in general unsteady flows.} 
\textcolor{black}{On the other hand, short-time attractors (Figs. \ref{fig:s1_vel_chick}b,d) are frame invariant, can move and deform, and apply to general unsteady flows.} Here we devise an optimal control framework that uses the active stress intensity $m$ as the control input to achieve a target short-time attractor, defined by a desired distribution of $s_1$ or iLE, in active viscous flows. %XX we can possibly add the fact that this is the first attempt to create/control active traps and list a few applications XX

\section{ \label{state} Active Fluid Model}
\begin{figure*}[t]
\includegraphics[width = .9\linewidth]{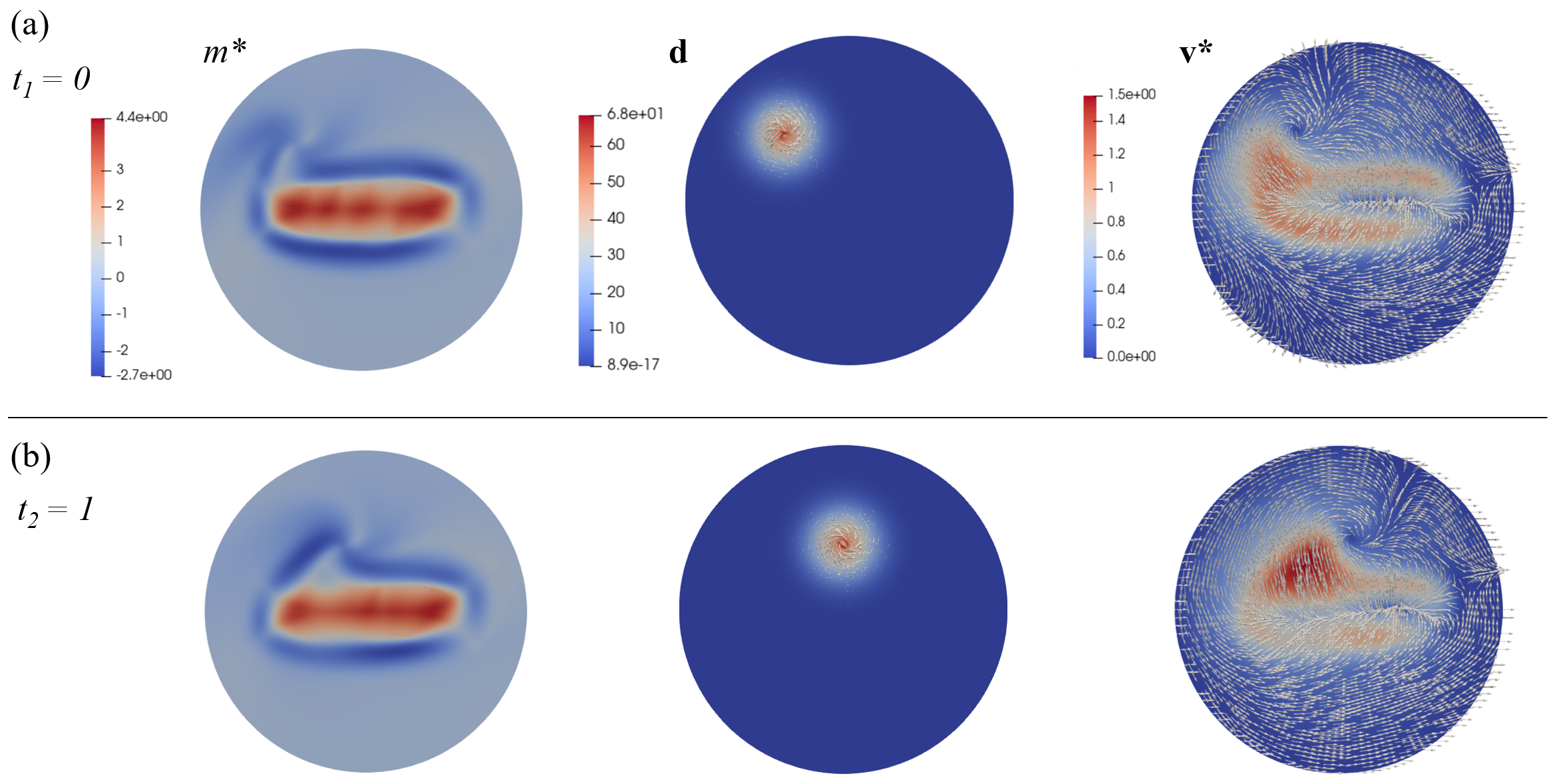}
\caption{
\label{mvd}
Optimal control state pair $(m^{\star},\mathbf{v}^{\star})$ and moving disturbance $\mathbf{d}$ associated to the OCP described in Fig. \ref{st_attr}. $\mathbf{d}$ and $\mathbf{v}^{\star}$ are vector fields (arrows) with their magnitude also displayed in the colormap. The $\mathbf{v}^{\star}$ vectors are normalized to ease visualization. Each row (a-b) corresponds to different initialization times $t_i$ as Fig. \ref{st_attr}. The velocity dynamics $\mathbf{v}$ are strongly affected by the presence of the disturbance. %The optimal control finds an optimal trade-off between counteracting and exploiting the disturbance to generate the short-time attractor shown in Figure \ref{st_attr}.
}
\end{figure*}

We adopt a simplified version of the mechanochemical model developed in \cite{serra2021mechanochemical} consisting of an active stokes flow characterized by the passive viscous stress $\mathbf{\sigma}_v = -p\mathbf{I} + 2\mu \mathbf{S}_d$ and active stress $\mathbf{\sigma}_a = m(\mathbf{B}-\mathbf{I}/2)$, where $\mathbf{S}_d$ is the deviatoric rate-of-strain tensor, $\mathbf{B} = \cv{e} \otimes \cv{e}$ characterizes the orientation of active elements $\cv{e} = (\cos(\phi) \, \sin(\phi))^{\top}$, $m$ denotes the intensity of active stress and $\mathbf{I}$ the identity tensor. To account for flow compressibility, we use a simple continuity equation $\mathbf{\nabla \cdot v} = c (-2p -p_0 m) $ where positive isotropic viscous stress ($p>0$), and isotropic contractile-type ($m>0$) active stress contribute to negative flow divergence via the bulk viscosity $1/c$ and a nondimensional parameter $p_0$. Biologically, $p_0$ modulates the cell propensity to ingress into the third dimension given active isotropic apical contraction. The resulting system of PDEs in nondimensional form \cite{serra2021mechanochemical} is 
\begin{equation*}
\begin{array}{l}
2 p_{1} \Delta \cv{v}+\nabla[\nabla \cdot \cv{v}]+\cv{g}(m,\phi)=\mathbf{0}, \\
\mathbf{g}(m,\phi) = p_1[2(\mathbf{B}\nabla m +m \nabla \cdot  \mathbf{B}) + (p_0 -1) \nabla m] = \nabla \cdot (\mathbf{A} m),\\
%\phi_{t}+(\cv{v} \cdot \nabla) \phi=\frac{\omega}{2}+p_{2}\left(\frac{u_{y}+v_{x}}{2} \cos 2 \phi+\frac{v_{y}-u_{x}}{2} \sin 2 \phi\right) \\
\phi_{t} = \mathbf{f_1(\cv{v},\phi)};\ m_{t} = \mathbf{f_2}(\cv{v},m),\\
%m_{t}+(\cv{v} \cdot \nabla) m=p_{3}-p_{4} m e^{-\frac{p_{5}}{2} m}+p_{6} \Delta m
\end{array}
\end{equation*}
where $p_1=\mu c$ is a second nondimensional parameter characterizing  the ratio of the shear to bulk viscosity, and $\bb{A}=2 p_1 \mathbf{B} + p_1(p_0-1) \mathbf{I}$. The first two terms in the force balance describe passive forces due to shear and compressibility, while the active forces  $\mathbf{g}$ arise from spatial variations of $m$ and $\phi$. To set up our optimal control problem, we consider a simplified dynamics where the orientation of active elements $\phi(\bb{x})$ is time-independent and prescribed, while the active stress intensity $m(\mathbf{x},t)$ is the control input. 
%\crl{Giustificare questa scelta, experiment design, attivazione fotosensibile etc}
%In this case, the model (or state dynamics) simplifyes to 
%\begin{equation}
%\label{eq:State}
%\begin{array}{ll}
%  -2p_1 \Delta \cv{\mathbf{v}}  -\nabla [\nabla \cdot  \cv{\mathbf{v}}] =  \cv{g}(m)+\cv{d}  & \textrm{in} \quad \Omega  \\
%  \cv{v}=\cv{0} & \textrm{on} \quad \partial \Omega.
% \end{array}
%\end{equation}
%where we have added a time-dependent disturbance $\cv{d}(\mathbf{x},t)$, and denoted by $\Omega$ a two-dimensional domain with boundary $\partial \Omega$ on which we enforce zero velocity. %Although the PDE is time-dependent due to the effect of $\mathbf{d}$, there are no explicit time derivative operators as we neglect inertia forces. %In this setting, time can be considered as a dependent and we can consider time as a parameter. 

%\begin{figure*}
%\centering 
%\includegraphics[width = 1\linewidth]{Figures/fig_vel.png}
%\caption{
%\label{asym_attr}
%Optimal solution of the linear quadratic OCP. The target velocity $\mathbf{z}=(-4x,-8y)$ is selected to generate a straight-line shaped asymptotic attractor. The optimal state velocity $\mathbf{v}^{\star}$ is shown at the leftmost plot while snapshots of particles advection is shown in the remaining plot. Streamlines are in black.}
%\end{figure*} 

\begin{comment}
As a consequence, the state equation has the same structure found in linear elasticity theory i.e. Navier-Cauchy equations which, however, describe the displacement dynamics. 
\end{comment}

\section{ \label{ocp} Results} 
%In the first case, we track a desired velocity profile $\mathbf{z}$, which has an asymptotic attractor such as a sink or a saddle-type fixed point. We formulate the OCP 
%as follows 
%\begin{equation}
%\label{OCP_v}
%\begin{aligned}
%   & \min_{m,\cv{v}} \,\,J_{\mathbf{v}} = \frac{1}{2} \int_{\Omega} (\cv{v} - \cv{z})^2\,d\Omega  + \frac{\beta}{2} \int_{\Omega} \Big( m^2 + \norm{\nabla m}^2 \Big) d\Omega   \\
%     & \vspace{0.5cm} \\
%     & s.t. \, \begin{array}{ll}
%\displaystyle -2p_1 \Delta \cv{v}  -\nabla [\nabla \cdot  \cv{v}] = \cv{g}(m) +\cv{d} &  \textrm{in} \quad \Omega  \\
%\vspace{-3mm} \\
% \cv{v} = \cv{0} & \textrm{on} \quad \partial \Omega, \\
%\end{array}
%\end{aligned}
%\end{equation}
%which is linear in the state dynamics since $\mathbf{g}(m) = \nabla \cdot (\mathbf{A} m)$ and quadratic in the cost functional. \textcolor{black}{In the presence of time-varying disturbances, however, there are generally no asymptotic attractors, while short-time attractors exist and shape material transport over limited time intervals \cite{serra2016objective,serra2020search}. 
To control short-time attractors, the OCP involves steering the minimum eigenvalue of the rate of strain tensor towards a target function while minimizing the overall control effort and its gradient:
\begin{equation}
\label{OCP_l}
\begin{aligned}
    & \min_{m,\cv{v}} \,\,J_{s} =\frac{1}{2} \int_{\Omega} (s_1 - z)^2\,d\Omega  + \frac{\beta}{2} \int_{\Omega} \Big( m^2 + \norm{\nabla m}^2 \Big) d\Omega,\\
     & \vspace{0.5cm} \\
     & s.t. \, \begin{array}{ll}
\displaystyle -2p_1 \Delta \cv{v}  -\nabla [\nabla \cdot  \cv{v}] = \cv{g}(m) +\cv{d} &  \textrm{in} \quad \Omega  \\
\vspace{-3mm} \\
 \cv{v} = \cv{0} & \textrm{on} \quad \partial \Omega, \\
\end{array}
\end{aligned}
\end{equation}
where $z(\mathbf{x})$ represents a scalar target for the minimum eigenvalue $s_1(\mathbf{x})$ of the rate-of-strain tensor $\mathbf{S}$ of $\cv{v}$, and $\cv{d}(\mathbf{x},t)$ is a disturbance on the state dynamics.  

Following a Lagrangian variational approach, we derive a system of first-order necessary optimality conditions for the OCP problem. We note that  $J_{s}$ is nonlinear and nonquadratic due to the relation $s_1(\mathbf{S}(\cv{v}))$. Consequently, sufficiency is not guaranteed, as typical in nonconvex problems. We define the Lagrangian as
\begin{eqnarray}
    && \mathcal{L} = J_{s}  +\int_{\Omega} \cs{\lambda} \cdot \Big(\cv{g}(m)+ \cv{d} + 2p_1 \Delta \cv{v}  +\nabla [\nabla \cdot  \cv{v}]\Big) \, d\Omega, \nonumber
\end{eqnarray}
where $\cs{\lambda}$ is the adjoint function, and obtain the strong form of the optimality conditions by imposing the first variations of the Lagrangian to be zero for all allowed variations of the state, adjoint and control functions. We provide the complete derivation in the SM Sections 2-4, and summarize here the necessary conditions as a coupled system of PDEs. 
%\noindent We use the notation $\mathcal{L}'(\cv{v})\cs{\psi}$ to denote the first variation (Gateaux derivative) of the cost functional $\mathcal{L}$ computed at $\cv{v}$ in the direction $\cs{\psi}$. 
The optimal pair for the OCP \eqref{OCP_l} $(\cv{v}, \, m)$ should satisfy the following system of first-order necessary conditions
\begin{equation}
\label{nc}
\begin{array}{ll}
    -2p_1 \Delta \cv{v}  -\nabla [\nabla \cdot  \cv{v}] = \cv{g}(m)+\cv{d} & \textrm{in} \quad \Omega\\
    \cv{v}=\cv{0}  & \textrm{on} \quad \partial\Omega \\
    %-2p_1 \Delta \cs{\lambda}  -\nabla [\nabla \cdot  \cs{\lambda}] =\cv{v}-\cv{z} & \textrm{in} \quad \Omega\\
     -2p_1 \Delta \cs{\lambda}  -\nabla [\nabla \cdot  \cs{\lambda}] =-\nabla \cdot \Big((s_1-z)\,\cs{\xi}_1 \otimes \cs{\xi}_1 \Big) & \textrm{in} \quad \Omega\\
    \cs{\lambda}=\cv{0}  & \textrm{on} \quad \partial\Omega \\
    -\beta \Delta m + \beta m - \nabla \cs{\lambda} : \mathbf{A} = 0 & \textrm{in} \quad \Omega,\\
\end{array}
\end{equation}
%while for the OCP \eqref{OCP_l} the adjoint equation is replaced by
%\begin{equation}
%\label{adj_nl}
%\begin{array}{ll}
%    -2p_1 \Delta \cs{\lambda}  -\nabla [\nabla \cdot  \cs{\lambda}] =-\nabla \cdot \Big((s_1-z)\,\cs{\xi}_1 \otimes \cs{\xi}_1 \Big) & \textrm{in} \quad \Omega,\\
%\end{array}
%\end{equation}
where $\cs{\xi}_1(\bb{x})$ is the eigenvector function associated to $s_1(\bb{x})$. We note that \eqref{nc} requires an iterative method due to the nonlinear relationship between $\cv{v}$ and the forcing term of the adjoint equation involving $s_1$ and $\cs{\xi}_1(\bb{x})$.

%\section{ \label{num} Results } 

\begin{figure*}[t]
\includegraphics[width = .9\linewidth]{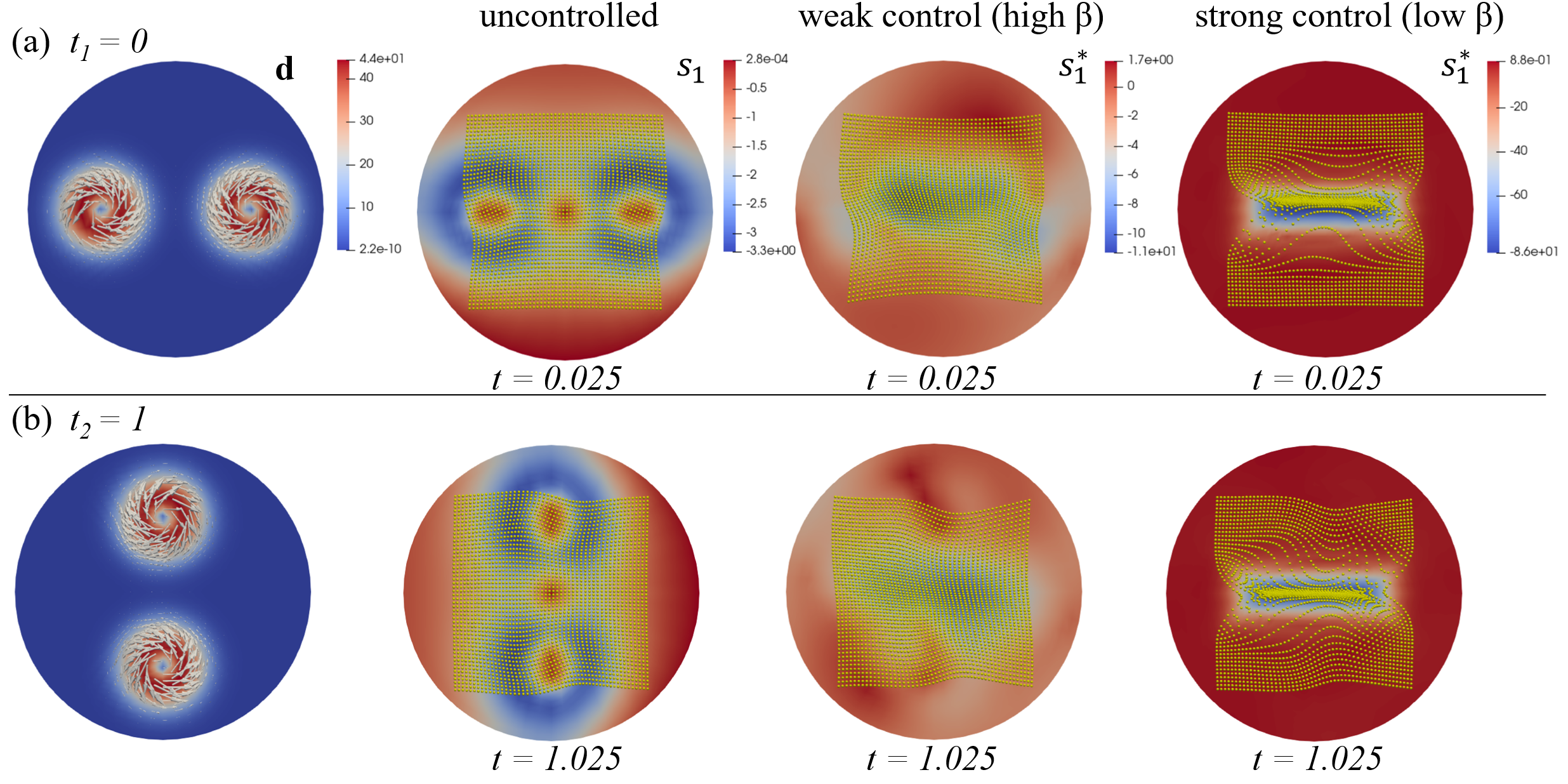}
\caption{
\label{large-short}
Controlled and uncontrolled dynamics for a different spatiotemporal disturbance $\mathbf{d}$ compared to Figs. \ref{st_attr}-\ref{mvd}. Each row (a-b) corresponds to a different time ($t_i,\ i=1,2$). The second column shows the $s_1(\mathbf{x})$ field in the case of no control along with fluid tracers (yellow dots) advected for a short time interval of length $0.025$ by the uncontrolled $\mathbf{v}(\mathbf{x},t)$, starting from a uniform initial grid at $t_i$. The third and fourth columns show the same as the second one in the case of weak ($\beta = 0.1$) and strong ($\beta = 10^{-6}$) control. Particles are advected under the optimal controlled velocity, whose corresponding $s_1^{\star}(\mathbf{x})$ field is in the colormap. A higher $\beta$ results in worse control performance due to its increasing relative importance in the cost functional. \textcolor{black}{Here, $\cv{d}(\mathbf{x},t) = \mathbf{d}(\bb{x}_{c1}(t)) - \mathbf{d}(\bb{x}_{c2}(t))$, where $\bb{x}_{c1}(t) = 0.5[\cos(0.5 \pi t + \pi), \cos(0.5 \pi t + \pi)],\ \bb{x}_{c2}(t) = 0.5[\cos(0.5 \pi t ), \cos(0.5 \pi t )]$, and $\mathbf{d}(\bb{x}_{ci}(t)) = [-(y-y_{ci}(t)),  (x-x_{ci}(t))] 500 \exp ( -((x-x_{ci}(t))^2+(y-y_{ci}(t)^2))/0.2^2$.}}
\end{figure*}

We numerically solve \eqref{nc} using the Finite Element Method (FEM) and a gradient-based algorithm (SM Sec. 5). We consider a circular domain with zero velocity boundary conditions and note that our algorithm applies to arbitrary-shaped domains. We set the target shape $z$ as a scaled indicator function of a rectangle so that the target value is $-10$ inside the rectangle and zero elsewhere. We set the cable orientation to a constant value $\phi=\frac{\pi}{4}$ from the x-axis and choose the control weighting parameter $\beta=10^{-6}$ and the nondimensional model parameters $p_0=10,\ p_1=0.5$. $p_1$ modulates the overall fluid compressibility while high $p_0$ induces high negative divergence in regions with higher $m$ \cite{serra2021mechanochemical}. We select the space-time varying disturbance force as \textcolor{black}{$\cv{d}(\mathbf{x},t) = d\,e^{-(\frac{r(t)}{\sigma})^2}\,[-(y-y_c(t)),x-x_c(t)]$, where $\cv{x}_c(t) = [-0.5+t,0.5]$, $r(t) = ||\cv{x}-\cv{x}_c(t)||$, and set the intensity $d=50$, and standard deviation $\sigma = 0.2$.} Figure \ref{st_attr} shows the resulting optimal $s_1$ along with a grid of particles advected over short times for two different initialization times. 
%\textcolor{green}{Carlo, puoi descrivere intuitivamente la format del disturbo e scrivre i parametric precise? tipo x0 e y0 (facendo rirerimento al sistema di coordinate in fig 2a? grazie) - simile per la figura 4}
\begin{comment}
\crl{
Figura 4: 
\begin{equation}
\mathbf{d}(\bb{x}_i(t)) = \begin{bmatrix}
- (y-y_i(t)) \\
   (x-x_i(t))
\end{bmatrix}
500 \exp ( -((x-x_1(t))^2+(y-y_1(t)^2))/0.2^2     )
\end{equation}

\begin{equation}
\mathbf{d}(t) = \mathbf{d}(\bb{x}_1(t)) - \mathbf{d}(\bb{x}_2(t))
\end{equation}

\begin{equation}
\bb{x}_1(t) = 0.5
\begin{bmatrix}
\cos(0.5 \pi t + \pi) \\
\cos(0.5 \pi t + \pi)
\end{bmatrix} \quad
\bb{x}_2(t) = 0.5
\begin{bmatrix}
\cos(0.5 \pi t) \\
\sin(0.5 \pi t)
\end{bmatrix} 
\end{equation}

Figura 3:
\begin{equation}
\mathbf{d}(t) = \mathbf{d}_0(\bb{x}_0(t))
\end{equation}
\begin{equation}
\bb{x}_0(t) 
\begin{bmatrix}
-0.5+0.5t \\
0.5
\end{bmatrix} 
\end{equation}
}
%The optimal control $\mathbf{m}^{\star}$ generates a velocity field that accurately tracks the target, which in turn generates an asymptotic attractor shown in Figure \ref{asym_attr}. The asymptotic attractor generation strategy through a linear quadratic OCP is effective and computationally efficient. 
\end{comment} 

%\begin{figure*}[h!]
%\includegraphics[width = 1 \linewidth]{Figures/fig_large_short.png}
%\caption{Moving disturbance in the large time, advection of particles in short time for uncontrolled low beta and high beta case $\beta=1,10^{-4}$}
%\end{figure*}

Figure \ref{mvd} shows the optimal state-control pair and its associated disturbance forcing $\cv{d}$. The control $m$ acts through $\mathbf{g}(m)=\nabla \cdot (\mathbf{A} m)$, and therefore both $m$ and $\mathbf{\nabla} m$ contribute to the state dynamics. This is why the control generates sharp gradients to track the target eigenvalue accurately. Furthermore, the control field rejects the disturbance by generating a vortex-shaped counteracting effect. Overall the disturbance strongly influences the optimal velocity. %, from which it's difficult to infer the presence of the desired short-time attractor region, which is evident from Figure \ref{st_attr}. %\textcolor{green}{Carlo, puoi metteere i lables a) prima riga, b) seconda riga e c) terza riga e aggiuntare il testo dove citiamo questi subpanels? Anche nelle altre figure se necesario, grazie. Sarebbe figo e sbalorditivo se l'attrattore che stiamo creando rimaga invisible alle streamlines. Ora nell'ultima riga, si intravede che il campo di velocita converge sull'asse x piu o meno. Se aggiungessimo a questo campo di velocita' una constant speed in direzione x, annulleremmo questo effetto ma l'attrattore rimarrebbe lo stesso. Oppure, se rescribessiamo questo campo di velocita in un sistema di coordinate che ruota con una certa velocita rispetto a quello attuale, le frecce del campo di velocita risulterebbero circolari ma l'attrattore rimarrebbe unaltered. Possiamo parlarne a voce.} 
Figure \ref{large-short} shows the interplay between the control weight $\beta$, the disturbance $\mathbf{d}$ and the accuracy of the tracking objective that is key to generating a short-time attractor. As in the earlier figures, each row (a-b) corresponds to a different time ($t_i,\ i=1,2$). 

To present an additional test case, we select a different disturbance compared to Figs. \ref{st_attr}-\ref{mvd}, generating two vortex-shaped force-field streamlines (Figure \ref{large-short}, first column) using the same functional form in the previous test case. The target eigenvalue $z$ is the same as in Figs. \ref{st_attr}-\ref{mvd}. The uncontrolled dynamics ($m=0$) does not generate attraction, as shown by short-term advected particles (yellow dots) and the $s_1(\mathbf{x})$ field in Figure \ref{large-short}, second column. A high $\beta=0.1$ (or weak control) steers the eigenvalue towards the target, but its effect is too mild to generate short-time attraction (Figure \ref{large-short}, third column). Indeed, a higher control weight would result in a lower relevance of the tracking term in the cost function and hence poor tracking performances. Figure \ref{large-short} fourth column shows the results for a low $\beta=10^{-6}$ (or strong control), as in Figure \ref{st_attr}, resulting in the desired short-time attractor that serves as a material trap while rejecting disturbances. Different rows (a-b) show the same results for two different initialization times as in Figs. \ref{st_attr}-\ref{mvd}.

\section{ \label{conc} Conclusion}

By combining recent advances in coherent structures from nonlinear dynamics, control theory and active matter, we formulated, analyzed and numerically solved an optimal control problem that generates material short-time attractors at desired locations while rejecting disturbances.
We used a simplified version of model \cite{serra2021mechanochemical} describing a compressible, viscosity-dominated, active fluid representing a planar multicellular flow, and used the active stress intensity as the control input. We identify short-time attractors as the smallest instantaneous Lyapunov exponent \cite{Nolan2020}, the instantaneous limit of the well-known FTLE, which is computable from the smallest eigenvalue of the rate of strain tensor of the flow velocity. 

Short-time attractors are frame invariant and predict the correct location of material attraction, which may be undetected from the inspection of the frame-dependent velocity field (Fig. \ref{fig:s1_vel_chick}, \cite{serra2016objective,serra2020search}). Using the same framework, one can control the position of material repellers, which, together with attractors, shape complex motion in synthetic active matter \cite{serra2021defect} and embryo morphogenesis across species \cite{serra2020dynamic, lange2023zebrahub}. Controlling attractors and repellers offers a new, robust, and frame-invariant perspective to steer motion in synthetic and natural active matter. It will enable the creation of material traps for medical applications as well as enhance our understanding and the possibility of manipulating multicellular flows in morphogenesis. For example, it will shed light on how myosin activity (active stress intensity) generates the required motion that compartmentalizes the embryo, segregating distinct cell types (repellers) and steering specific cells to precise locations (attractors). 
In future work, we plan to consider the explicit orientational dynamics of the active stress anisotropy, the effect of inertial forces, and the control of Lagrangian Coherent Structures that shape fluid motion over longer time scales.

\begin{acknowledgments}
We acknowledge Alex Plum for his comments on the manuscript. 
\end{acknowledgments}

\bibliography{apssamp,Biblio_MS}% Produces the bibliography via BibTeX.
\clearpage

\preprint{APS/123-QED}

\title{Supplementary material: Optimal Control of Short-Time Attractors in Active Fluid Flows}% Force line breaks with \\

\author{Carlo Sinigaglia}
%\email{carlo.sinigaglia@polimi.it}
\author{Francesco Braghin}%
% \email{francesco.braghin@polimi.it}
\affiliation{%
 Politecnico di Milano, Department of Mechanical Engineering, 20156, Italy
}%


\author{Mattia Serra}
 \email{mserra@ucsd.edu.}
\affiliation{University of California San Diego, Department of Physics, CA 92093, USA
}%
%\affiliation{
%\affiliation{%


\date{\today}% It is always \today, today,

\maketitle

%\tableofcontents

%\input{Sections/SM}

%\setcounter{figure}{0}
%\setcounter{section}{0}
%\setcounter{equation}{0}
\renewcommand{\thefigure}{S\arabic{figure}}

\renewcommand{\thefigure}{S\arabic{figure}}
\renewcommand{\theequation}{S\arabic{equation}}
\renewcommand{\thesection}{S\arabic{section}}

\section{S1. Notation}
We adopt a slightly different notation than the main text to distinguish continuous functions and their associated numerical discretization. Continuous space-varying functions are denoted with lowercase bold arrowed symbols (i.e. $\cv{v} \in H_0^1(\Omega)^2$) while their numerical discretization as the standard bold vector $\bb{v} \in \mathbb{R}^{N}$ where $N$ is the number of degrees of freedom of the Finite Element discretization. Furthermore, matrices arising from the numerical discretization are denoted with uppercase letters i.e. $A \in \mathbb{R}^{N \times N}$.

\section{S2. Analysis of the state dynamics}

In order to set up the problem in the proper setting we need some basic notions from functional analysis, see e.g. \cite{salsa2015partial}. We denote $L^2(\Omega)^2$ the space of vector-valued square integrable functions, that is $\{ \cv{f} | \int_{\Omega} \norm{\cv{f}}^2 d\Omega < \infty \}$ then, $H_0^1(\Omega)^2$ denotes the space of vector-valued square integrable functions with zero trace and square integrable gradient , that is $\{ \cv{f} \in L^2(\Omega)^2 \, | \, \int_{\Omega} \norm{\nabla \cv{f}}_{F}^2 < \infty \, \text{and} \, \cv{f}_{\partial \Omega} = \cv{0} \}$ while $H^{-1}(\Omega)^2$ denotes its dual, that is the space of continuous linear functionals defined on $H_0^1(\Omega)^2$. We can now select the appropriate spaces for our problem. The state dynamics $\cv{v}\in H_0^1(\Omega)^2$ while the control input $m \in H^1(\Omega)$. In this setting, the control operator can be defined as the linear operator $\mathbf{g}:H^1(\Omega) \mapsto H^{-1}(\Omega)^{2}$. The function $\cv{g}(m) \in L^2(\Omega)^2$ since $m \in H^1(\Omega)$, as we will prove in the following theorem on the well-posedness of the state dynamics. The weak formulation of the state dynamics reads: find $\cv{v} \in H_0^1(\Omega)^2$ such that
\begin{equation*}
    \int_{\Omega} \Big(-2p_1 \Delta \cv{v}  -\nabla [\nabla \cdot  \cv{v}]\Big) \cdot \cs{\phi}\, d\Omega = \int_{\Omega} \cv{g}(m) \cdot \cs{\phi}\, d\Omega  \, 
\end{equation*}
$\forall \cs{\phi} \in H_0^1(\Omega)^2$, which can be rearranged in a more convenient form using integration by parts and homogeneous boundary conditions as

\begin{equation*}
    \int_{\Omega} 2p_1 \nabla \cv{v} : \nabla  \cs{\phi} + (\nabla \cdot \cv{v}) (\nabla \cdot  \cs{\phi})\, d\Omega = \int_{\Omega} -m \, \mathbf{A} : \nabla \cs{\phi}\, d\Omega 
\end{equation*}
$\forall \cs{\phi} \in H_0^1(\Omega)^2 \, $, where the operator $:$ denotes the Frobenius innner product of two matrices, that is $\displaystyle \mathbf{A}:\mathbf{B}=\sum_{i}^n\sum_{j}^n A_{ij}B_{ij}$. Note that the Frobenius inner product induces a norm in the space of matrices that we denote as $\norm{\mathbf{A}}_F := \sqrt{\mathbf{A}:\mathbf{A}}$, for more details, see e.g. \cite{horn2012matrix}. Existence and uniqueness of solutions to this problem follow from an application of the Lax-Milgram lemma of functional analysis, see e.g. \cite{salsa2015partial}. In particular, we can set the state problem in abstract form by defining the symmetric bilinear form $a(\cv{v},\cs{\phi}) :=  \int_{\Omega} 2p_1 \nabla \cv{v} : \nabla  \cs{\phi} + (\nabla \cdot \cv{v}) (\nabla \cdot  \cs{\phi})\, d\Omega$ and the linear functional $F_m\cs{\phi}:= \int_{\Omega} \cv{g}(m) \cdot \cs{\phi}d\Omega $. In this setting, we can prove the following result.

\begin{proposition}
There exists a unique weak solution $\cv{v}$ to 
\begin{equation*}
    a(\cv{v},\cs{\phi}) = F_m \cs{\phi} \quad \forall \cs{\phi} \in H_0^1(\Omega)^2.
\end{equation*}
Furthermore, $\cv{v}$ depends continuously on the control function $m$ that is
\begin{equation*}
    \norm{\cv{v}}_{H_0^1(\Omega)^2} \leq \frac{\sqrt{\alpha}C_p}{2 p_1} \norm{m}_{H^1(\Omega)^2}
\end{equation*}
where $C_p$ is the Poincaré constant depending only on $\Omega$ and $\alpha = \max \{ \norm{\nabla \cdot \mathbf{A}}_{L^{\infty}(\Omega)}^2,\norm{\mathbf{A}}_{L^{\infty}(\Omega)}^2\}$ does not depend on $\cv{v}$ and $m$.
\begin{proof}
The control operator $\cv{g}$ that appears on the right-hand side of the state equation can be bounded with respect to the $H^1$-norm of the control function $m$ as
\begin{eqnarray*}
    &&\norm{\cv{g}}_{L^2(\Omega)^2}^2 \leq \norm{(\nabla \cdot \mathbf{A}) m}^2_{L^2(\Omega)^2} + \norm{\mathbf{A} \nabla m}^2_{L^2(\Omega)^2} \nonumber \\ 
    &&\leq \norm{\nabla \cdot \mathbf{A}}_{L^{\infty}(\Omega)}^2 \norm{m}_{L^2(\Omega)}^2 + \norm{\mathbf{A}}_{L^{\infty}(\Omega)}^2 \norm{\nabla m}_{L^2(\Omega)}^2 \\ 
    &&\leq \alpha \norm{m}^2_{H^1(\Omega)^2} \nonumber
\end{eqnarray*}
where $\alpha = \max \{ \norm{\nabla \cdot \mathbf{A}}_{L^{\infty}(\Omega)}^2,M\}$. Note that we have used the triangle inequality and the operator bound on the matrix $\mathbf{A}$ which is assumed to be sufficiently smooth. Once this bound is established, we can prove existence and uniqueness of a weak solution to the state equation by showing that the hypotheses of Lax-Milgram lemma are satisfied, that is the bilinear form is continuous and coercive and the linear functional $F_m$ is continuous with respect to the $H_0^1(\Omega)^2$-norm. Coerciveness can by proven as follows
\begin{eqnarray}
    &&a(\cv{v},\cv{v}) = \int_{\Omega} 2p_1 \nabla \cv{v} : \nabla  \cv{v} + (\nabla \cdot \cv{v}) (\nabla \cdot  \cv{v})\, d\Omega \nonumber \\
    &&\geq \int_{\Omega} 2p_1 \nabla \cv{v} : \nabla  \cv{v} d\Omega = 2 p_1 \norm{\cv{v}}^2_{H_0^1(\Omega)^2} \nonumber 
\end{eqnarray}
since, thanks to Poincaré inequality, $\sqrt{\int_{\Omega} \nabla \cv{v} : \nabla \cv{v} \, d\Omega}$ is a norm in $H_0^1(\Omega)^2$. Continuity of the linear functional can be inferred by 
\begin{equation*}
    \begin{aligned}
    & |F_m \cs{\psi}| \leq \norm{\cs{g}(m)}_{L^2(\Omega)^2}\norm{\cs{\psi}}_{L^2(\Omega)^2} \\ 
    &\leq \sqrt{\alpha} \,C_p \, \norm{m}_{H^1(\Omega)^2} \norm{\cs{\psi}}_{H_0^1(\Omega)^2}
    \end{aligned}
\end{equation*}
using the bounds of the control operator $\cv{g}$ with respect to the norm of $m$ and the Poincarè inequality, i.e. $\norm{\cs{\psi}}_{L^2(\Omega)^2} \leq C_p \norm{\nabla \cs{\psi}}_{L^2(\Omega)^2} = C_p \norm{\cs{\psi}}_{H_0^1(\Omega)^2}$. Continuity of the state dynamics $\cv{v}$ with respect to the control $m$ is proven by plugging in $\cv{v}$ as test function in the weak formulation of the problem and using the coercivity of the bilinear form. Indeed we have
\begin{equation*}
\begin{aligned}
    &2 p_1 \norm{\cv{v}}^2_{H_0^1(\Omega)^2} \leq a(\cv{v},\cv{v}) = F_m \cv{v}  \\
    &\leq \sqrt{\alpha}C_p \norm{m}_{H^1(\Omega)^2} \norm{\cv{v}}_{H_0^1(\Omega)^2}
\end{aligned}
\end{equation*}
from which we get
\begin{equation*}
    \norm{\cv{v}}_{H_0^1(\Omega)^2} \leq \frac{\sqrt{\alpha}C_p}{2 p_1} \norm{m}_{H^1(\Omega)^2}.
\end{equation*} \QEDA
\end{proof}

\end{proposition}

%\crl{Da questo dovrebbe arrivare anche la differenziabilità della mappa stato-controllo data la linearità del problema}.

\section{S3. Analysis of the Optimal Control Problem}

 We have already established existence and uniqueness of a solution to the state equation. It is left to prove that the functional $J$ is weakly sequentially lower semicontinous.  The term of the cost functional involving the control $m$ is trivially weakly lower semicontinuous since it is the $H_1(\Omega)$-norm of the control function squared. We will now prove that the first term involving the eigenvalue $s_1$ is continuous in $\cv{v}$ and thus weakly sequentially lower semicontinuous. %\crl{(Trovare il teorema che dimostra che se è continua allora è anche weakly sequentially lower semicontinua.)} 

Before deriving a system of first-order necessary conditions for optimality we prove well-posedness by showing that $s_1 \in L^2(\Omega)$, indeed
\begin{eqnarray}
    && \norm{s_1}_{L^2(\Omega)}^2 = \int_{\Omega}s_1^2 d\Omega \leq \int_{\Omega}(s_1^2 + s_2^2) d\Omega \nonumber \\
    && \leq \int_{\Omega} \norm{\mathbf{S}(\cv{v})}_{F}^2 d\Omega \leq \int_{\Omega} \norm{\nabla \cv{v}}^2_{F} d\Omega = \norm{\cv{v}}_{H_0^1(\Omega)^2}^2
\end{eqnarray}
thanks to the inequality $\sum_{i=1}^{n} s_i^2 \leq \norm{\mathbf{A}}_{F}^2$ involving the eigenvalues of a matrix $\mathbf{A}$. The second inequality comes from the following reasoning, it holds that:
\begin{equation*}
    2 \bb{S}(\cv{v}):\bb{S}(\cv{v}) - \nabla \cv{v} : \nabla \cv{v} = \nabla \cdot ( \nabla \cv{v} \,\cv{v} - (\nabla \cdot \cv{v})\cv{v}) + (\nabla \cdot \cv{v})^2
\end{equation*}
Integrating over $\Omega$ and using the boundary conditions we get
\begin{eqnarray}
    && \int_{\Omega} \norm{\mathbf{S}(\cv{v})}_{F}^2 d\Omega = \frac{1}{2}\Big( \int_{\Omega} \norm{\nabla \cv{v}}^2_{F} d\Omega +   \int_{\Omega}(\nabla \cdot \cv{v})^2 d\Omega \Big) \nonumber \\ 
    && \leq \int_{\Omega} \norm{\nabla \cv{v}}^2_{F} d\Omega
\end{eqnarray}
since $(\nabla \cdot \cv{v})^2 = (\text{Tr}(\nabla \cv{v}))^2 \leq \norm{\nabla \cv{v}}_F^2$.

We denote $J_{s}[\cv{v}] := \int_{\Omega} (s_1-z)^2 d\Omega$ and show that $|J_{s}[\cv{v}_n]-J_{s}[\cv{v}]| \to 0$ as $\cv{v}_n \to \cv{v}$. We also denote $s_{1,n}:=s_1(\cv{v}_n)$ and $s_{1}:=s_1(\cv{v})$ to ease the notation. Note also that $J_{s}[\cv{v}] = \int_{\Omega} (s_1-z)^2 d\Omega = (s_1-z,s_1-z)_{L^2(\Omega)} = \norm{s_1-z}^2_{L^2(\Omega)}$ can be seen either as a scalar product in $L^2(\Omega)$ or as a tracking type cost functional involving the norm of $s_1-z$, which is well-posed since $s_1 \in L^2(\Omega)$. Indeed, we can write
\begin{eqnarray}
    && J_{s}[\cv{v}_n]-J_{s}[\cv{v}] = (s_{1,n}-z,s_{1,n}-z)_{L^2(\Omega)} \\
    && - (s_1-z,s_1-z)_{L^2(\Omega)} \nonumber \\
    && = (s_{1,n},s_{1,n})_{L^2(\Omega)}-(s_1,s_1)_{L^2(\Omega)}\nonumber  \\
    && - 2(s_{1,n}-s_1,z)_{L^2(\Omega)} \pm (s_{1,n},s_1)_{L^2(\Omega)} \nonumber \\
    && = (s_{1,n},s_{1,n}-s_1)_{L^2(\Omega)} + (s_1,s_{1,n}-s_1) - 2(s_{1,n}-s_1,z) \nonumber \\
    && \leq \norm{s_{1,n}}_{L^2(\Omega)}\norm{s_{1,n}-s_1}_{L^2(\Omega)} \\
    && + \norm{s_{1}}_{L^2(\Omega)}\norm{s_{1,n}-s_1}_{L^2(\Omega)} \nonumber \\ 
    && + 2\norm{z}_{L^2(\Omega)}\norm{s_{1,n}-s_1}_{L^2(\Omega)} \nonumber \\
    && = \Big( \norm{s_{1,n}}_{L^2(\Omega)} + \norm{s_{1}}_{L^2(\Omega)} + 2\norm{z}_{L^2(\Omega)} \Big)\norm{s_{1,n}-s_1}_{L^2(\Omega)} \nonumber \\
    && = C \norm{s_{1,n}-s_1}_{L^2(\Omega)} \nonumber
\end{eqnarray}

where we have used the properties of the inner product and the Cauchy-Schwarz inequality in $L^2(\Omega)$. Note that the constant $C:=\Big( \norm{s_{1,n}} + \norm{s_{1}} + 2\norm{z} \Big)$ is well-defined and finite since $s_{1,n},s_1,z \in L^2(\Omega)$. Now, if we can prove that $\norm{s_{1,n}-s_1}_{L^2(\Omega)} \leq L \norm{\cv{v}_n-\cv{v}}_{H_0^1(\Omega)^2}$ for some constant $L>0$,the result follows. In proving that this indeed holds we need a theorem from eigenvalues perturbation analysis from \cite{horn2012matrix}, Chapter 6, Corollary 6.3.8 that we state here. If $\mathbf{A}$ and $\mathbf{E}$ are symmetric matrices, denoting $\hat{s}$ the eigenvalues of $\mathbf{A}+\mathbf{E}$ and by $s$ the eigenvalues of $\mathbf{A}$ it holds that 
$
    \sum_{i=1}^{n} |\hat{s}_i - s_i|^2 \leq \norm{\mathbf{E}}_F^2
$. We can write
\begin{eqnarray}
&& \norm{s_{1,n}-s_1}^2_{L^2(\Omega)}  = \int_{\Omega} (s_{1,n}-s_1)^2 d\Omega  \\
&& \leq \int_{\Omega} \norm{\mathbf{S}(\cv{v}_n)-\mathbf{S}(\cv{v})}_F^2 d\Omega \nonumber = \int_{\Omega} \norm{\mathbf{S}(\cv{v}_n-\cv{v})}_F^2 d\Omega \nonumber \\
&& \leq \int_{\Omega} \norm{\nabla(\cv{v}_n-\cv{v})}_F^2 d\Omega = \norm{\cv{v}_n-\cv{v}}^2_{H_0^1(\Omega)^2} \nonumber
\end{eqnarray}
where in applying Corollary 6.3.8 we set $\mathbf{A} = \mathbf{S}(\cv{v}(\mathbf{x}))$ and $\mathbf{E} = \mathbf{S}(\cv{v}_n(\mathbf{x}))-\mathbf{S}(\cv{v}(\mathbf{x}))$ which are both symmetric.

\begin{comment}
An existence result on solutions to the OCP \eqref{OCP} is provided by the following theorem 
\begin{theorem}
Consider the minimization problem \eqref{OCP} subject to the state dynamics where the control function $m \in H^1(\Omega)$. Then, there exists a pair $(\cv{v},m)$ which minimizes $J$.
\begin{proof}
See the SM. \QEDA
\end{proof}
\end{theorem}
\end{comment}

\section{S4. Derivation of the optimality system}

\noindent The Gateaux derivative of $\mathcal{L}$ computed at $\cv{v}$ in the direction $\cs{\psi}$ can be written as:
\begin{eqnarray}
\label{L'v}
    && \mathcal{L}'(\cv{v})\cs{\psi}  = \frac{\partial }{\partial \epsilon}\Big\rvert_{\epsilon=0} \Big(\frac{1}{2} \int_{\Omega} (s_1(\mathbf{S}(\cv{v}+\epsilon \cs{\psi})) - z)^2\,d\Omega \\ 
    && + \int_{\Omega} \cs{\lambda} \cdot \Big(2p_1 \Delta (\cv{v}+\epsilon \cs{\psi})  +\nabla [\nabla \cdot  (\cv{v}+\epsilon \cs{\psi})]\Big) \, d\Omega\Big).\nonumber
\end{eqnarray}
In the following we will need the differentiation formula for a non repeated eigenvalue $\lambda$ with respect to a symmetric matrix function $\mathbf{A}(\epsilon)$ parametrized with respect to a scalar $\epsilon \in \mathbb{R}$, see e.g. \cite{horn2012matrix}, Theorem 6.3.12, Chapter 6. Indeed, we have $ \displaystyle
    \frac{d \lambda }{d \epsilon}\Big\rvert_{\epsilon=\epsilon_0} = \cs{\xi}(\epsilon_0)^{\top}\frac{d \mathbf{A}}{d \epsilon}\Big\rvert_{\epsilon=\epsilon_0} \cs{\xi}(\epsilon_0)
$
where $\cs{\xi}$ is the associated normalized eigenvector.

The first term in \eqref{L'v} can be rearranged using integration by parts and the linearity of Frobenius inner product as
\begin{eqnarray}
     && \frac{\partial }{\partial \epsilon}\Big\rvert_{\epsilon=0} \Big(\frac{1}{2} \int_{\Omega} (s_1(\mathbf{S}(\cv{v}+\epsilon \cs{\psi})) - z)^2\,d\Omega \nonumber \\
     && =  \int_{\Omega} (s_1(\mathbf{S}(\cv{v})) - z) \frac{\partial }{\partial \epsilon}\Big\rvert_{\epsilon=0} s_1(\mathbf{S}(\cv{v}+\epsilon \cs{\psi}) \,d\Omega \nonumber \\
     &&= \int_{\Omega} (s_1-z)\,\cs{\xi}_1^{\top} \cv{S}(\cs{\psi})\cs{\xi}_1 d\Omega  
     = \int_{\Omega} (s_1-z)\,\cs{\xi}_1\otimes \cs{\xi}_1 : \mathbf{S}(\cs{\psi})  d\Omega \nonumber \\ 
     &&= - \int_{\Omega} \nabla \cdot \Big((s_1-z)\,\cs{\xi}_1\otimes \cs{\xi}_1\Big) \cdot \cs{\psi}  d\Omega  \nonumber \\
     && + \int_{\Omega} \nabla \cdot ( (s_1-z)\,\cs{\xi}_1\otimes \cs{\xi}_1 \cs{\psi} ) d\Omega \nonumber \\
     &&= - \int_{\Omega} \nabla \cdot \Big((s_1-z)\,\cs{\xi}_1\otimes \cs{\xi}_1\Big) \cdot \cs{\psi}  d\Omega \nonumber \\
     && +\cancel{ \int_{\partial \Omega}  ( (s_1-z)\,\cs{\xi}_1\otimes \cs{\xi}_1 \cs{\psi} )d\Omega} \nonumber
\end{eqnarray}
where we have used the fact that for a matrix-valued function $\mathbf{A}(\mathbf{x})$ and a vector-valued function $\cs{\psi}(\mathbf{x})$ it holds that $
    \nabla \cdot (\mathbf{A} \cs{\psi}) = (\nabla \cdot \mathbf{A}) \cdot \cs{\psi} + \mathbf{A}:\nabla \cs{\psi}.
$
Using integration by parts, the second term of Equation \eqref{L'v} reads:
\begin{eqnarray}
    && \frac{\partial }{\partial \epsilon}\Big\rvert_{\epsilon=0}\int_{\Omega} \cs{\lambda} \cdot \Big(2p_1 \Delta (\cv{v}+\epsilon \cs{\psi})  +\nabla [\nabla \cdot  (\cv{v}+\epsilon \cs{\psi})]\Big) \, d\Omega\Big) \nonumber \\ 
    && = \int_{\Omega}  \Big(2p_1 \Delta \cs{\lambda}  +\nabla [\nabla \cdot  \cs{\lambda}]\Big) \cdot \cs{\psi}\, d\Omega. \nonumber
\end{eqnarray}

\noindent From the optimality condition 
\begin{equation*}
    \mathcal{L}'(\cv{v})\cs{\psi} = 0 \quad \forall \cs{\psi} \in H_0^1(\Omega)
\end{equation*}
we obtain the strong form of the adjoint equation which is a linear elliptic PDE of the form
\begin{equation*}
\begin{array}{ll}
    -2p_1 \Delta \cs{\lambda}  -\nabla [\nabla \cdot  \mathbf{\cs{\lambda}}] = - \nabla \cdot \Big((s_1-z)\,\cs{\xi}_1\otimes \cs{\xi}_1 \Big) & \textrm{in} \quad \Omega\\
    \cs{\lambda}=\cv{0}  & \textrm{on} \quad \partial\Omega
\end{array}
\end{equation*}
where $\cs{\xi}_1(\bb{x})$ is the space varying eigenvector associated to the minimum eigenvalue $s_1(\bb{x})$ and $\cs{\lambda}(\bb{x})$ is the adjoint variable.
\noindent The Euler equation can be recovered by taking the first variation of the Lagrangian with respect to the control function $m$.
\begin{equation*}
    \mathcal{L}'(m)h = \beta \Big(\int_{\Omega} m\,h + \nabla m \cdot \nabla h  \, d\Omega \Big) + \frac{\partial }{\partial \epsilon}\Big\rvert_{0} \int_{\Omega} \cs{\lambda}\cdot \cv{g}(m+\epsilon h) \, d\Omega.
\end{equation*}
Due to the linearity of the inner product and of $\cv{g}$ we have:
\begin{eqnarray}
    && \frac{\partial }{\partial \epsilon}\Big\rvert_{0} \int_{\Omega} \cs{\lambda}\cdot \cv{g}(m+\epsilon h) \, d\Omega = \frac{\partial }{\partial \epsilon}\Big\rvert_{0} \int_{\Omega} \bs{\lambda}\cdot \Big(\cv{g}(m) +\epsilon \cv{g}(h)\Big) \, d\Omega \nonumber \\
    && = \int_{\Omega} \cs{\lambda} \cdot \cv{g}(h) d\Omega. \nonumber 
\end{eqnarray}
Using integration by parts and the explicit form of $\cv{g}$ we get:
\begin{equation*}
    \int_{\Omega} \cs{\lambda} \cdot \cv{g}(h) d\Omega = \int_{\Omega} \cs{\lambda} \cdot \nabla \cdot\Big(\mathbf{A} h \Big)  d\Omega = -\int_{\Omega}  \nabla \cs{\lambda} : \mathbf{A} \, h  \, d\Omega 
\end{equation*}
since $\cs{\lambda}$ should satisfy homogeneous Dirichlet boundary conditions.
The Euler equation is obtained by the optimality condition:
\begin{equation*}
    \mathcal{L}'(m)h = 0 \quad \forall h \in H^1(\Omega)
\end{equation*}
which gives %\crl{pensare se servono boundary su $m$ quando è in $H^1(\Omega)$}
\begin{equation*}
    -\beta \Delta m + \beta m - \nabla \cs{\lambda} : \mathbf{A} = 0.
\end{equation*}

\begin{figure*}[t!]
\centering 
\includegraphics[width = .8\linewidth]{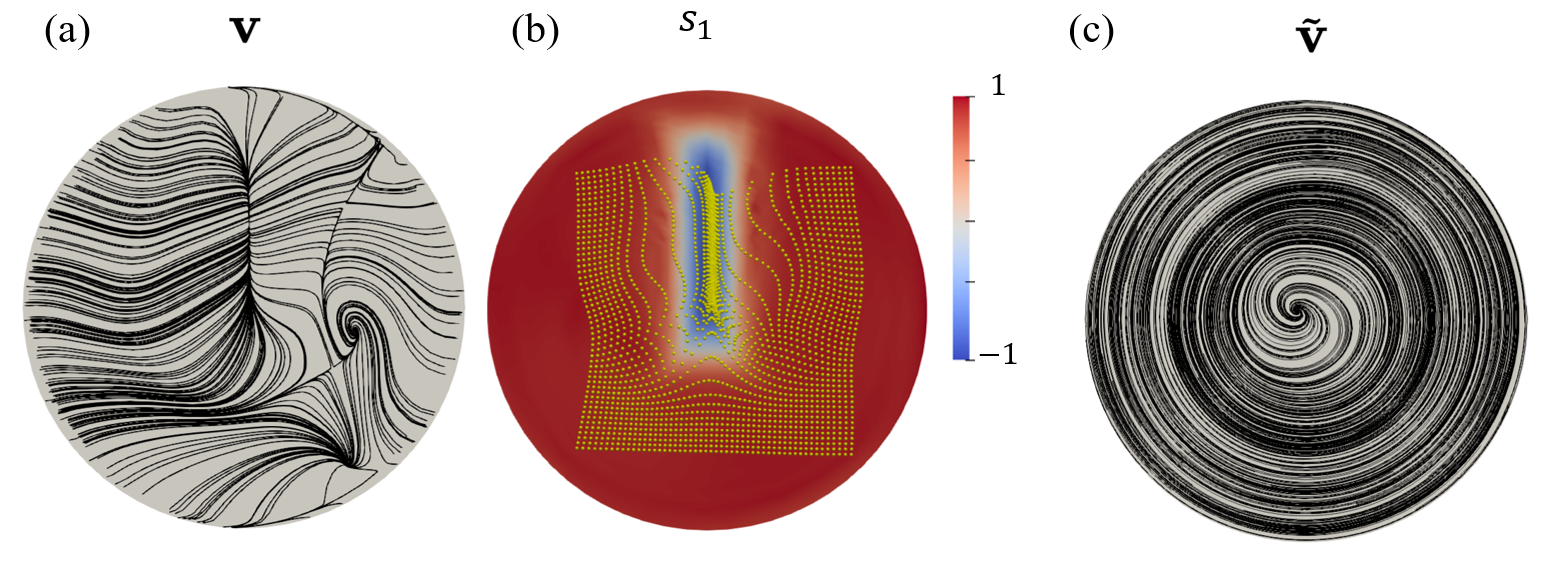}
\caption{
\label{tilde}
Solution of the OCP in eqs. (1-2) from the main text, setting $\mathbf{d}=\mathbf{0}$ and $z_1$ to create a short-term attractor along the positive $y-$axis. (a) Streamlines of the velocity field $\bold{v}$. (b) $s_1(\mathbf{x})$ marks short-time attraction, confirmed by fluid tracers advected by $\bold{v}$ over a short time, as Figs. 1,2,4. (c) Streamlines of the velocity field $\tbold{v}$. The tilde frame is rotating with respect to the standard frame at constant angular velocity $\omega=120\ rad $ (time is in nondimensional units). \textcolor{black}{The rotated frame is displayed at the time when it is instantaneously aligned with the standard frame. While $\bold{v}$ and the streamlines are frame-dependent, $s_1(\mathbf{x})=\tilde{s}_1(\mathbf{\tilde{x}})$ is not, correctly predicting short-term attractors regardless of the arbitrary choice of the reference frame used to describe motion.}}
%is invariant with respect to the time-varying rotation. }
\end{figure*} 

\section{S5. Details on the numerical implementation}
We select linear piecewise continuous finite elements and use the software package \texttt{deal.II} \cite{dealII93} for the finite element approximation while \texttt{NLopt} \cite{Johnson2011} is used for the gradient-based optimization. The discretized system of optimality conditions read
\begin{equation*}
\begin{array}{l}
A \bb{v} = B \bb{m} \\
A \boldsymbol{\lambda} = \mathbf{f}(\mathbf{v}) \\
\beta (A_m+M) \bb{m} + B^{\top} \boldsymbol{\lambda} = \mathbf{0}\\
\end{array}
\end{equation*}
where $A$ and $M$ are the standard stiffness and mass matrix of the Finite Element formulation of vector-valued and scalar-valued elliptic problems, respectively. The discretized input operator is $B_{ij} = - \int_{\Omega} \nabla \cs{\psi}_i : \mathbf{A} \, \psi_j\, d\Omega$ where $\cs{\psi}_i$ and $\psi_j$ are the associated vector and scalar basis functions, respectively. The nonlinear forcing in the adjoint equation is obtained from the weak formulation of the adjoint equation as $\mathbf{f}_i = \int_{\Omega} (s_1 - z)\, \cs{\xi}_1 \otimes \cs{\xi}_1 : \mathbf{S}(\cs{\psi}_i) d\Omega$ where $s_1(\bb{v})$ and $\cs{\xi}_1(\bb{v})$ are obtained from the numerical solution of $\bb{v}$. At each iteration of the gradient-based optimizer the reduced gradient can be computed by solving state and adjoint equation, then we have $\nabla J = \beta (A_m+M) \bb{m} + B^{\top} \boldsymbol{\lambda}$ which is used for the control update. Where not otherwise stated, the optimization is completed when the $L^2$ norm of the reduced gradient is less than $10^{-7}$.

\section{S6. Objectivity}
\begin{figure*}
\centering 
\includegraphics[width = 1\linewidth]{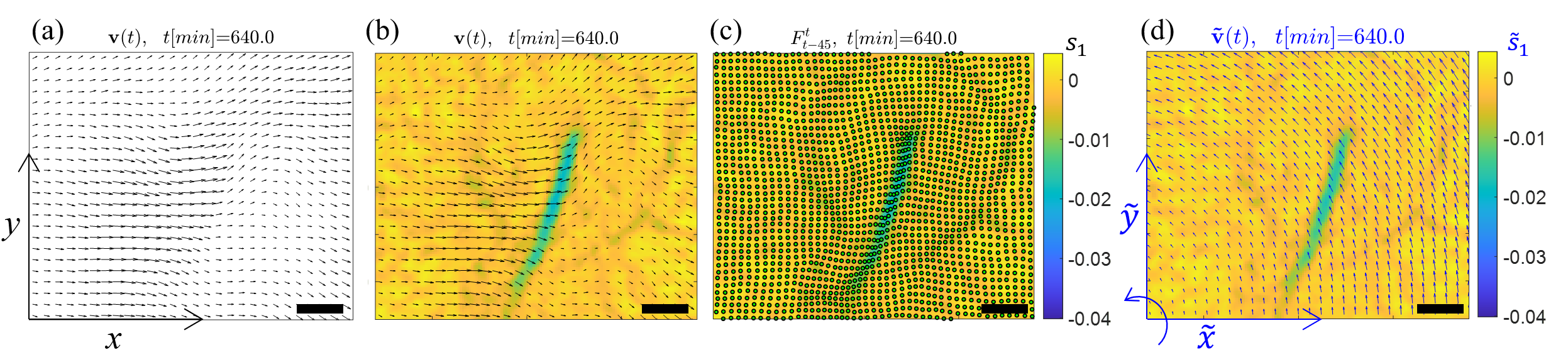}
\caption{
\label{fig:s1_vel_chickve}
\textcolor{black}{Same as Figs. 1a-b for a different time. (a) Velocity field $\mathbf{v}(\mathbf{x})$. (b) $\mathbf{v}(\mathbf{x})$ and $s_1(\mathbf{x})$ field. (c) $s_1(\mathbf{x})$ locates a short-term attractor, as verified by fluid tracers (green) advected by $\mathbf{v}(\mathbf{x})$ over a short time from 
 uniform initial conditions, as in Figs. 1b. (d) Velocity field $\tilde{\mathbf{v}}(\tilde{\mathbf{x}})$ obtained from eqs. (\ref{eq:v vtilda},\ref{eq:Q}), setting $\omega=1\ rad/min$, $\dot{\mathbf{b}}=\mathbf{0}$ and starting with $\tilde{\mathbf{x}}$ and $\mathbf{x}$ aligned at the time shown $t=640$. $\mathbf{v}(\mathbf{x})$ and its streamlines are frame dependent and hence are strongly affected by the choice (motion) of the reference frame. By contrast, $s_1(\mathbf{x})=\tilde{s}_1(\mathbf{\tilde{x}})$ is objective and predicts the correct location of short-term attractors.}}
\end{figure*} 

Assume $\mathbf{v}(\mathbf{x},t)$ is the velocity field solving the OCP (eq. (2)) or one experimentally measured as in Figs. 1a-b. Objectivity is a fundamental axiom of mechanics \cite{gurt} that states that the material response of a deforming continuum is independent of the reference frame chosen to describe the motion, for all frames related by Euclidean (or distance preserving) transformations of the form: $\tilde{\mathbf{x}}=\mathbf{Q}(t)\mathbf{x} + \mathbf{b}(t)$, where $\mathbf{Q}(t) \in SO(2)$ and $\mathbf{b}(t)$ is a translation vector. Therefore, the location of short-time attractors, which is related to fluid deformation and motion, cannot depend on the reference frame choice. 

Denoting by $\tilde{(\cdot)}$ and $(\cdot)$ the same quantity expressed in the $\tilde{\mathbf{x}}$ and $\mathbf{x}$ frames, scalar objective quantities must transform \cite{gurt} as $\tilde{c}(\tilde{\mathbf{x}},t)=c(\mathbf{x},t)$, objective vector fields must transform as $\tilde{\mathbf{c}}(\tilde{\mathbf{x}},t)=\mathbf{Q}(t)\mathbf{c}(\mathbf{x},t)$, etc.. Short-time attractors are objective because $\tilde{s}_1(\tilde{\mathbf{x}},t)=s_1(\mathbf{x},t)$ \cite{serra2016objective}. By contrast, it is well known that the velocity field and its streamlines are not as
\begin{equation}
\tilde{\mathbf{v}}(\tilde{\mathbf{x}},t) = \mathbf{Q}(t) \mathbf{v}(\mathbf{Q}^{\top}\tilde{\mathbf{x}},t) +  \dot{\mathbf{Q}}(t)\mathbf{Q}^{\top}(t)\tilde{\mathbf{x}} + \textcolor{black}{\dot{\mathbf{b}}}(t).
\label{eq:v vtilda}
\end{equation}

\textcolor{black}{To visualize the effects of frame invariance,  we solve the OCP in eqs. (1-2) from the main text, setting for simplicity $\mathbf{d}=\mathbf{0}$, and $z_1$ in order to create a short-term attractor along the positive $y-$axis (Figs. \ref{tilde}a-b). We then select a coordinate frame $\tilde{\mathbf{x}}$ with $\dot{\mathbf{b}}=\mathbf{0}$ and constant angular velocity $\omega$ so that}
\begin{equation}
\mathbf{Q}(t) = \begin{bmatrix}
\cos(\omega t) & -\sin(\omega t) \\
\sin(\omega t) &  \cos(\omega t) 
\end{bmatrix}, 
\quad
\dot{\mathbf{Q}}(t) = \omega \begin{bmatrix}
-\sin(\omega t) & -\cos(\omega t) \\
 \cos(\omega t) & -\sin(\omega t)
\end{bmatrix}.
\label{eq:Q}
\end{equation}

The frame rotation does not affect $s_1$ and the location of material attraction, but distorts the streamlines of $\tilde{\mathbf{v}}$ as shown in Figure \ref{tilde}c, highlighting the fact that the streamlines or the velocity field are suboptimal to identify material attraction. \textcolor{black}{Using the same coordinate transformation, a similar conclusion holds using the experimental velocity field used in Fig.1, as shown in Fig. \ref{fig:s1_vel_chickve}.}


%\bibliography{apssamp,Biblio_MS}% Produces the bibliography via BibTeX.
\end{document}